\documentclass[floatfix,twocolumn,showpacs,amsmath,amssymb,superscriptaddress]{revtex4}

\usepackage{graphicx}% Include figure files
\usepackage{dcolumn}% Align table columns on decimal point
\usepackage{bm}% bold math
\usepackage{times}% use standard PostScript fonts

\newcommand{\BINP}{Budker Institute of Nuclear Physics, 630090 Novosibirsk, Russia}
\newcommand{\TPU}{Institute of Nuclear Physics at Tomsk Polytechnic University, 634050 Tomsk, Russia}
\newcommand{\LIYAF}{Saint-Petersburg Institute of Nuclear Physics, 188350 Gatchina, Russia}
\newcommand{\ANL}{Argonne National Laboratory, Argonne, IL 60439, USA}
\newcommand{\NIKHEF}{NIKHEF, P.O.Box 41882, 1009 DB Amsterdam, The Netherlands}
\newcommand{\RUTGERS}{Rutgers University, Piscataway, NJ 08855, USA}

\newcommand{\dewid}{0.7\linewidth}
\newcommand{\thwid}{0.20\linewidth}
\newcommand{\egwid}{0.45\linewidth}

\newcommand{\fdir}{figs}

\begin{document}
\title{
Measurement of tensor analyzing powers in deuteron photodisintegration.
}

\author{I.A.Rachek}
\affiliation{\BINP}
%\thanks{corresponding author, e-mail:I.A.Rachek@inp.nsk.su}
\author{L.M.Barkov}
\affiliation{\BINP}
\author{S.L.Belostotsky}
\affiliation{\LIYAF}
\author{V.F.Dmitriev}
\affiliation{\BINP}
\author{M.V.Dyug}
\affiliation{\BINP}
\author{R.Gilman}
\affiliation{\RUTGERS}
\author{R.J. Holt}
\affiliation{\ANL}
\author{B.A.Lazarenko}
\affiliation{\BINP}
\author{S.I.Mishnev} 
\affiliation{\BINP}
\author{V.V.Nelyubin}
\affiliation{\LIYAF}
\author{D.M.Nikolenko} 
\affiliation{\BINP}
\author{A.V.Osipov}
\affiliation{\TPU}
\author{D.H. Potterveld}
\affiliation{\ANL}
\author{R.S\lowercase{h}.Sadykov} 
\affiliation{\BINP}
\author{Y\lowercase{u}.V.Shestakov} 
\affiliation{\BINP}
\author{V.N.Stibunov}
\affiliation{\TPU}
\author{D.K.Toporkov} 
\affiliation{\BINP}
\author{H.\lowercase{de} Vries}
\affiliation{\NIKHEF}
\author{S.A.Zevakov} 
\affiliation{\BINP}

\date{\today}
\begin{abstract}

A new accurate measurement of the tensor analyzing powers 
T$_{20}$, T$_{21}$ and T$_{22}$ in deuteron photodisintegration 
has been performed.
Wide-aperture non-magnetic detectors allowed broad kinematic 
coverage in a single setup: 
$E_\gamma$ = 25 to 600 MeV, and $\theta_p^{cm}$ = 24$^\circ$ to 48$^\circ$ 
and 70$^\circ$ to 102$^\circ$.
The new data provide a significant improvement over the few existing measurements.
The angular dependency of the tensor asymmetries in deuteron photodisintegration
is extracted for the first time.

\end{abstract}

%\keywords{Deuteron Photodisintegration, Tensor analyzing powers}

\maketitle

%\section{Introduction}
The simplest nucleus, the deuteron, is a natural laboratory for the
study of nuclear forces. One of the most fundamental processes on the deuteron
is two-body photodisintegration (PD) $\gamma + {\rm d} \rightarrow p + n$. It has
been a subject of intensive experimental and theoretical  research for
over 70 years (see Ref.~\cite{GG} for a comprehensive review). 
However several important observables still are 
measured with insufficient accuracy or not measured at all. The
tensor analyzing powers accessible through measurement of target asymmetries 
in PD of tensor polarized deuteron are among the most poorly 
known. Polarization observables are expected to be sensitive to 
important dynamical details and thus allow in general much more stringent tests 
of theoretical models.
The tensor polarizations are especially interesting because
there is a correlation between the degree of tensor polarization and the spatial
alignment of the deuteron.  For example, spatial alignment of the target
deuterons can lead to large asymmetries from final state interactions in
PD.

A general expression for the cross-section of the two-body 
PD of the polarized deuteron is written as follows:
\begin{equation}
\label{csect}
\begin{split}
\frac{d\sigma}{d\Omega} 
 &=  \frac{d\sigma_0}{d\Omega} \left\{
     1-\sqrt{3/4}\ P_z\sin \theta_H \sin \phi_H\,\rm{T}_{11} \right. \\
 &\phantom{=-}\left. + \sqrt{1/2}P_{zz}\left[
 (3/2\cos^2 \theta_H - 1/2) \rm{T}_{20}\right.\right. \\
 &\phantom{= +\sqrt{1/2}P_{zz}}\left.\left. - \sqrt{3/8}\sin 2\theta_H\,\cos\phi_H\,\rm{T}_{21} \right.\right. \\
 &\phantom{= +\sqrt{1/2}P_{zz}}\left.\left. + \sqrt{3/8}\sin^2\theta_H\,
\cos 2\phi_H\,\rm{T}_{22} \right] \right\},
\end{split}
\end{equation}
with $\sigma_0$ the unpolarized cross-section,
$P_z$ ($P_{zz}$) the degree of vector (tensor) polarization of the target,
$\theta_H$ the angle between polarization axis and momentum of 
$\gamma$-quantum, and
$\phi_H$ the angle between the polarization plane 
(containing the polarization axis and momentum of the photon) 
and the reaction plane (containing momenta of the proton and neutron).
The tensor analyzing powers T$_{2I}$ are functions of
photon energy $E_\gamma$ and proton emission angle $\theta_p^{cm}$.

Only three measurements of tensor polarization observables in deuteron 
PD have been reported prior to this experiment \cite{vepp2,bonn,phaseI}.
Here we present the results of a new measurement of the tensor analyzing power 
components T$_{20}$, T$_{21}$ and T$_{22}$.

%\section{Experiment setup}
The measurements were performed at the 2 GeV electron storage ring VEPP--3.
A thin-wall open-ended storage cell fed by polarized deuterium gas from an 
Atomic Beam Source (ABS) was used as an internal target \cite{ABS}.
The ABS provides a polarized 
deuterium gas jet with an intensity of up to $8\times 10^{16}$ atoms/s 
and a very high degree of tensor polarization -- above 98\% from extreme 
values ($+1$ or $-2$), 
while vector polarization was always close to zero ($|P_z| < 0.02$). 
Polarization of the gas stored inside the cell is degraded due to
various depolarizing processes.
The target polarization was determined by the ``Low--Q'' polarimeter \cite{LOWQ}. 
Tensor polarization averaged over the whole run of data taking was found 
to be 
$P^+_{zz} = 0.341 \pm 0.025 \pm 0.011 $, 
where the first uncertainty is statistical and the second one is systematic. 
The ratio $P^-_{zz}/P^+_{zz}=-1.70\pm 0.15$ is obtained by the analysis of 
the data collected on polarized and unpolarized deuterium target.  

In this experiment we measured the counting rate asymmetry for
two signs of tensor polarization of the deuterium target in disintegration
of the deuteron by a 2-GeV electron scattered forward, i.e. at an angle 
$\vartheta_e\approx 0^\circ$. In final state we detect the proton and neutron
in coincidence, the scattered electron is not detected. 
Scattering of an electron at $0^\circ$ is equivalent to the radiation of 
a real photon, which is then absorbed by a deuteron -- that is why in such 
a set-up it is the {\em photo--}disintegration, that is studied.

During the run the polarization settings were switched 
every 30 seconds to suppress systematic uncertainties.
It takes less than a second to alternate the polarization
by changing a resonance magnetic field in the rf-transition unit, 
located inside the ABS.
The experimental tensor target asymmetry is defined as 
\begin{equation}
\label{asym}
%a^T = \sqrt{2}\frac{N^+-N^-}{P^+_{zz} N^--P^-_{zz} N^+},
a^T = \sqrt{2}(N^+-N^-)/(P^+_{zz} N^--P^-_{zz} N^+),
\end{equation}
where $N^+$ ($N^-$) denotes the number of events detected with positive
(negative) target polarization.

In order to disentangle the three components of tensor analyzing power we 
collected data for three settings of the magnetic field, 
defining the polarization axis:  
$\theta_{\rm{H0}} = 0^\circ$, 
$\theta_{\rm{H1}} = 54.7^\circ$ and 
$\theta_{\rm{H2}} = 125.3^\circ$~,
with $\phi_H$ close to $0^\circ$ 
for all settings. 
According to Eq.~\ref{csect} the target asymmetry
in these cases is proportional to \ $a^T_0\sim c_0\rm{T}_{20}$, 
\ $a^T_1\sim (-c_1{\rm T}_{21} + c_2{\rm T}_{22})$ and 
\ $a^T_2\sim (+c_1{\rm T}_{21} + c_2{\rm T}_{22})$ respectively.
Here $c_0, c_1, c_2$ are constants defined by a geometry of detector
and target.
Therefore all three tensor moments are separated unambiguously. 
Note that a term with ${\rm T}_{11}$ is suppressed in all three configurations
even for non-zero vector polarization $P_z$ due to its $\sin{\phi}$-dependence, 
see Eq.\ref{csect}.

\begin{figure}[!tb]
\begin{center}
\includegraphics[width=\dewid]{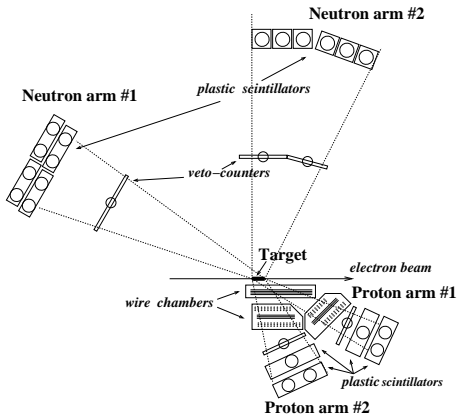}
\end{center}
\caption[]{
\label{detector}
Schematic side view of the particle detectors. 
}
\end{figure}

The particle detector consists of two pairs of arms for detecting protons and
neutrons in coincidence, as shown in Fig.~\ref{detector}. 
The first (second) pair covers an angle range of 
$\theta_p^{cm}=24^\circ$ to $48^\circ$ 
($\theta_p^{cm}=70^\circ$ to $102^\circ$).
Each proton arm includes wire chambers for tracking and 3 layers of plastic
scintillators ($2+12+12$cm thick). Neutron arms consist of plastic 
scintillators: 2-cm charged particle veto counters followed by 20-cm or $12+12$cm
scintillator bars placed at a largest available distance from 
the target, about 3 m, for the best TOF measurement. 
Azimuthal angular acceptance was $\Delta \varphi = 20^\circ$ 
for all arms.

\begin{figure}[!tb]
\begin{center}
\includegraphics[width=\dewid]{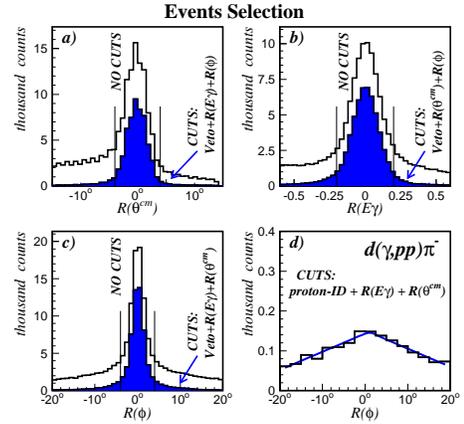}
\end{center}
\caption[]{
\label{ev_sel}
%(color online) 
Selection of two-body deuteron PD events. 
Panels a)--c) are histograms of correlation parameters: 
\ $R(\vartheta^{cm})=\vartheta^{cm}_{(p)}-\vartheta^{cm}_{(n)}-\pi$,
\ $R(E_\gamma)=(E_\gamma^{(p)}-E_\gamma^{(n)})/
	((E_\gamma^{(p)}+E_\gamma^{(n)})/2)$,
\ $R(\varphi)=\varphi_{p}-\varphi_{n}-\pi$. Shaded histograms are
for all cuts applied except the cut on the shown parameter. Vertical 
lines indicate the cut on the shown parameter.
Panel d) 
shows a shape of out-of-plane events distribution for the three-body
deuteron disintegration, which is selected by identifying a proton in the
neutron arm using TOF/E and $\Delta$E/E analysis. 
}
\end{figure}

%\section{Event selection}
Particle identification was based on the veto signal and TOF analysis for neutrons 
and on TOF and $\Delta$E/E analysis for protons. 
Further event selection relies on kinematic
correlations characteristic of the two-body PD -- see 
Fig~\ref{ev_sel}{\it a--c}. 
One such property is a coplanarity of proton and neutron momenta 
($|\phi _p - \phi _n| = \pi$).
Assuming deuteron two-body disintegration by a photon emitted along the
electron beam direction,
one can reconstruct $E_\gamma$ and cm angles from the momentum
vector of a single detected proton or neutron. This provides two more 
selection criteria: $E^{(p)}_{\gamma}=E^{(n)}_\gamma$ and
$\vartheta_{(p)}^{cm}+\vartheta_{(n)}^{cm}=\pi$. 
Such cuts allow both to reject the background and to constrain 
the electron scattering angle to $\vartheta_e < 0.5^\circ\ldots 2.0^\circ$,
depending on $E_\gamma$.
An amount of inseparable residual background, which comes mostly from
three-body PD  $d(\gamma, pn)\pi^0$, was estimated from the
analysis of a tail in the out-of-plane events distribution. The shape of
the tail is determined by selecting the events of a similar 3-body disintegration
process $d(\gamma, pp)\pi^-$ -- see Fig~\ref{ev_sel}{\it d}.  
The background analysis was performed separately for each polarization state.
After applying all cuts the fraction of unseparated background events 
was estimated %found 
to be from 1.5\% (low E$_\gamma$ region) to 5.6\%  (high E$_\gamma$ region).
The uncertainty from the unseparated background events is included in
the statistical uncertainty.

%\section{Systematic Errors}
The main source of systematic uncertainty is the uncertainty in the target 
polarization. The degree of polarization enters as a common factor for 
all data points. Other systematic uncertainties come from the inaccuracy 
of reconstruction of $E_\gamma$ and proton CM--emission angle. 
Contribution of these parameters
dominates at small $E_\gamma$ where tensor analyzing powers change fast with energy.
The false asymmetry related to fluctuations of other experimental parameters,
such as electron beam lifetime, target density variations, PMT gain instability,
{\it etc} is negligible because these fluctuations were completely unsynchronized with
the reversing of polarization, and characteristic time of fluctuations 
was much higher than the period of reversing of polarization (30 s).
Moreover, the beam current integral and time spent in each polarization state 
were measured precisely and taken into account in the analysis.  

\begin{figure}[!tb]
\begin{center}
\includegraphics[width=\egwid,clip=true]{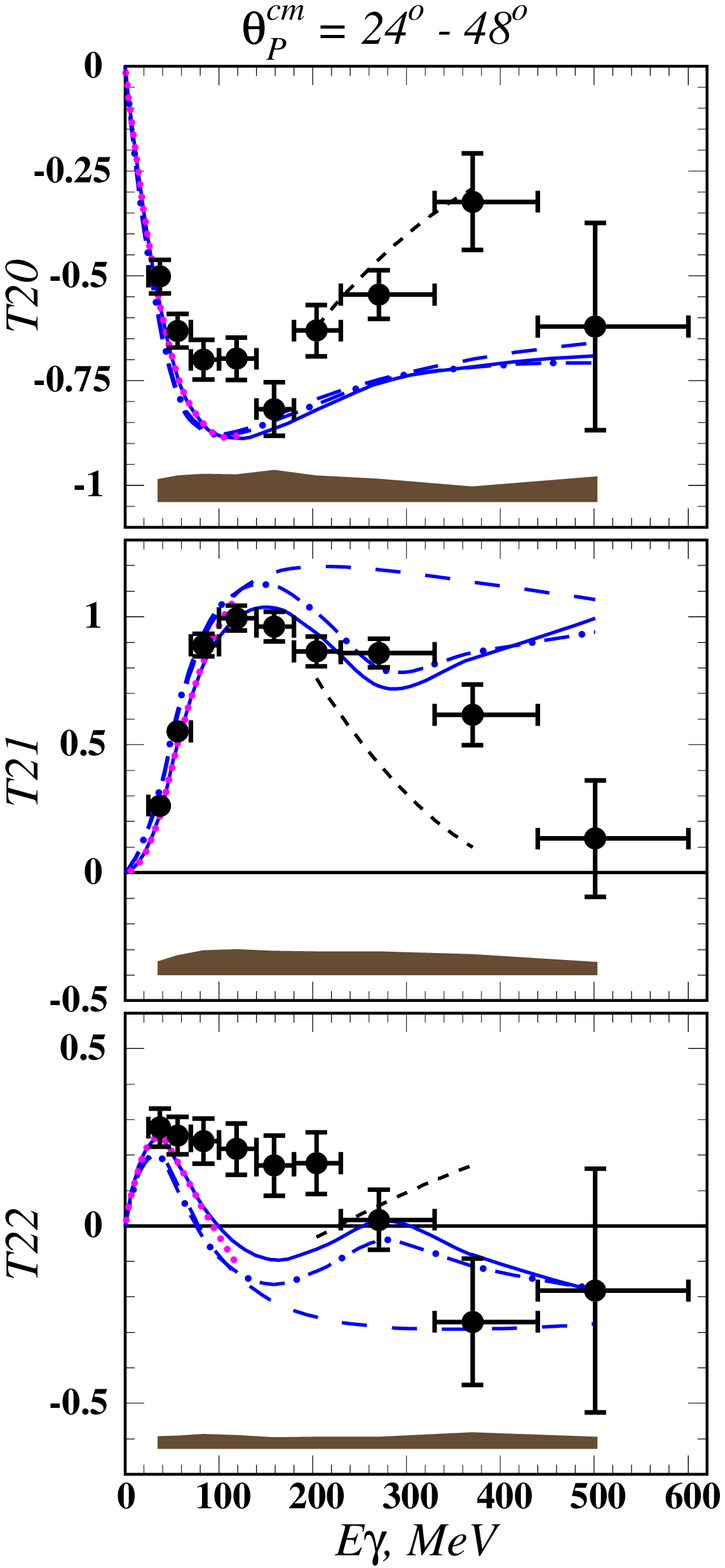}
\hspace*{0.05\linewidth}
\includegraphics[width=\egwid,clip=true]{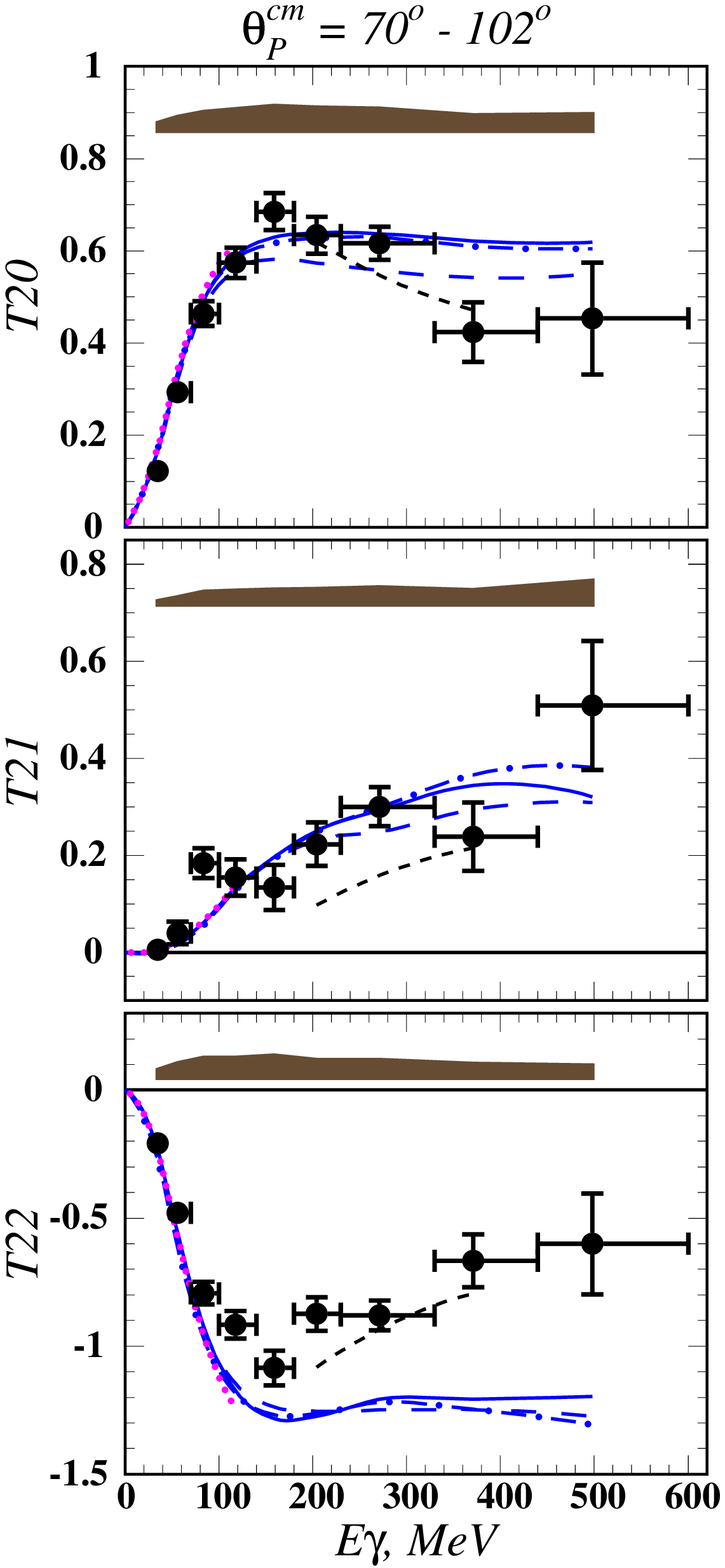}
\end{center}
\caption{
\label{egamma}(color online).
Tensor analyzing powers vs. photon energy. 
Vertical bars are statistical uncertainties; 
horizontal bars indicate the bin size. 
Shaded bands show systematic uncertainties.
Theoretical predictions are from
Arenh\"{o}vel \cite{arenho} 
"N+MEC" (blue long-dashed line), 
"N+MEC+IC" (blue dash-dotted line), and 
"N+MEC+IC+RC" (solid line) models,
from Levchuk \cite{levchuk} (magenta dotted line), and 
from Schwamb \cite{schwamb} (black short-dashed line).
}
\end{figure}

\begin{figure*}[hbt]
\begin{center}
\includegraphics[width=\thwid]{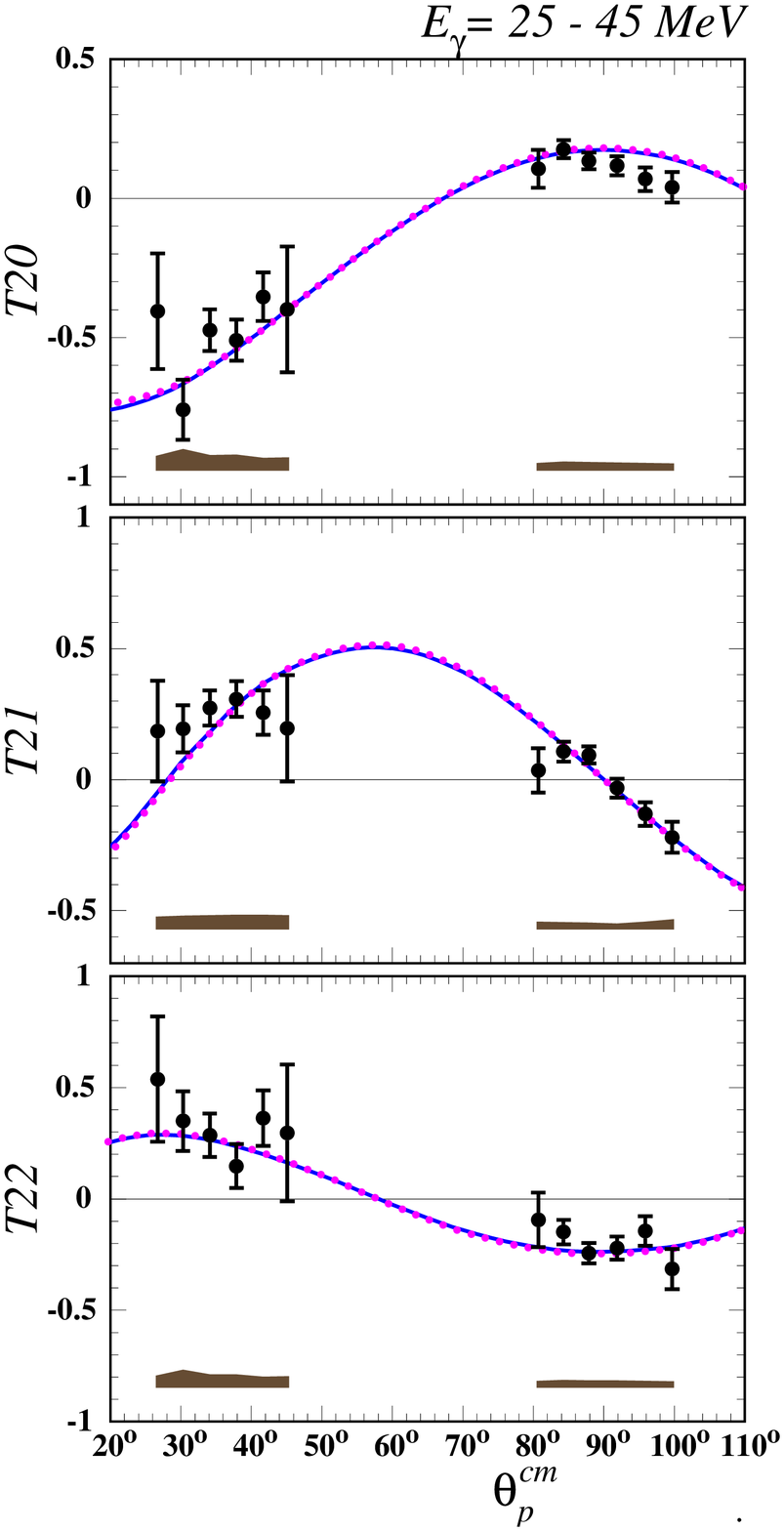}\hspace{0.01\linewidth}
\includegraphics[width=\thwid]{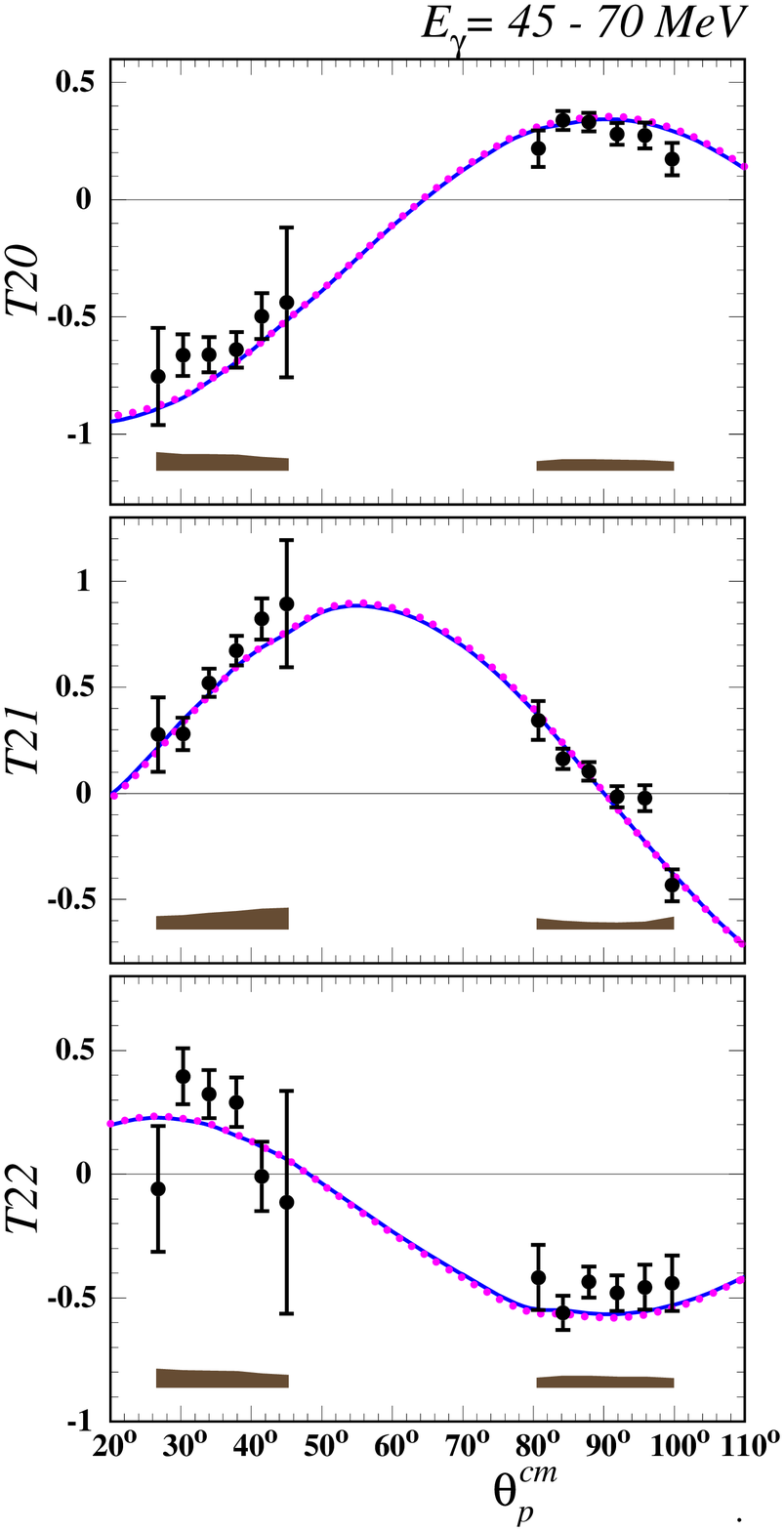}\hspace{0.01\linewidth}
\includegraphics[width=\thwid]{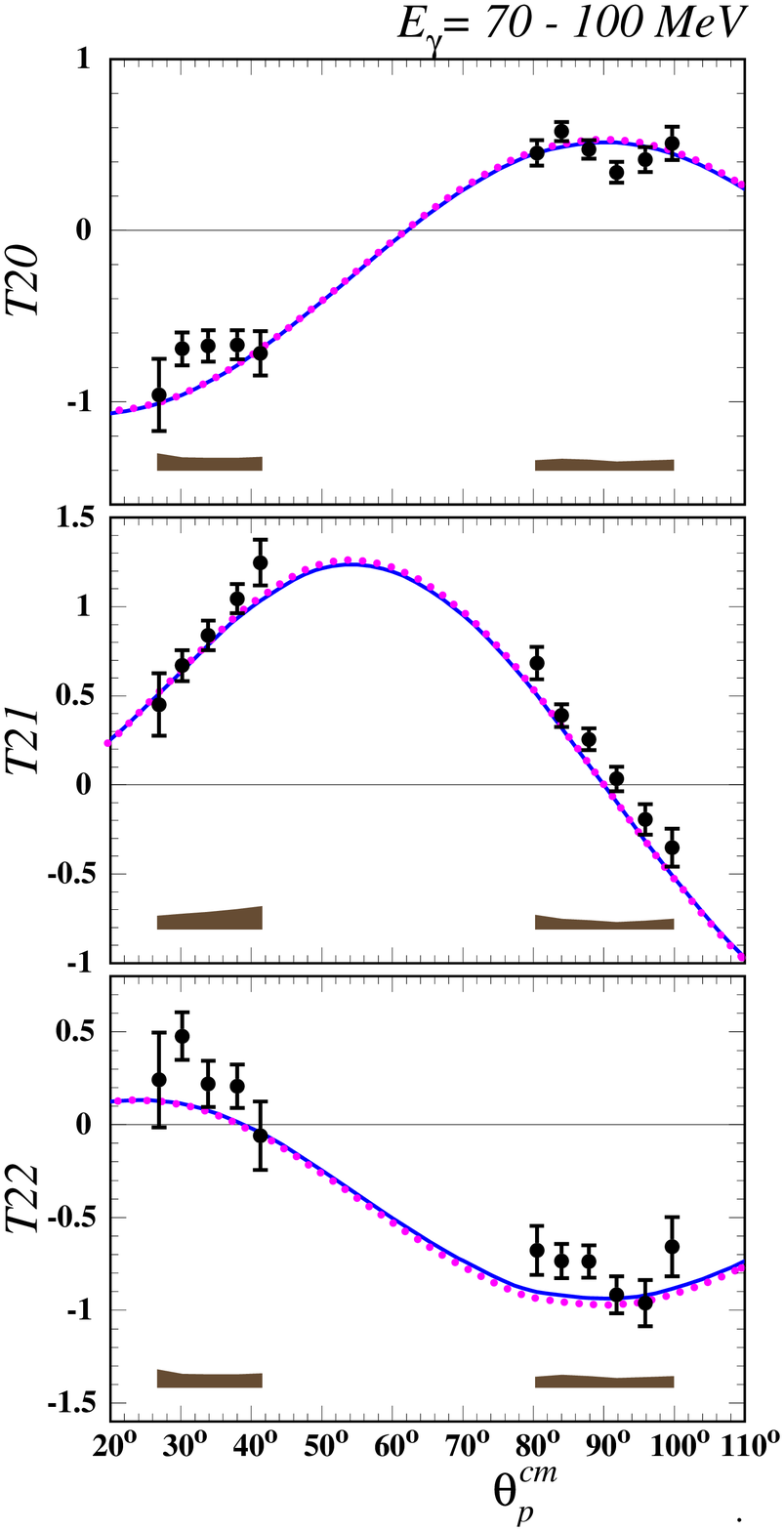}\hspace{0.01\linewidth}
\includegraphics[width=\thwid]{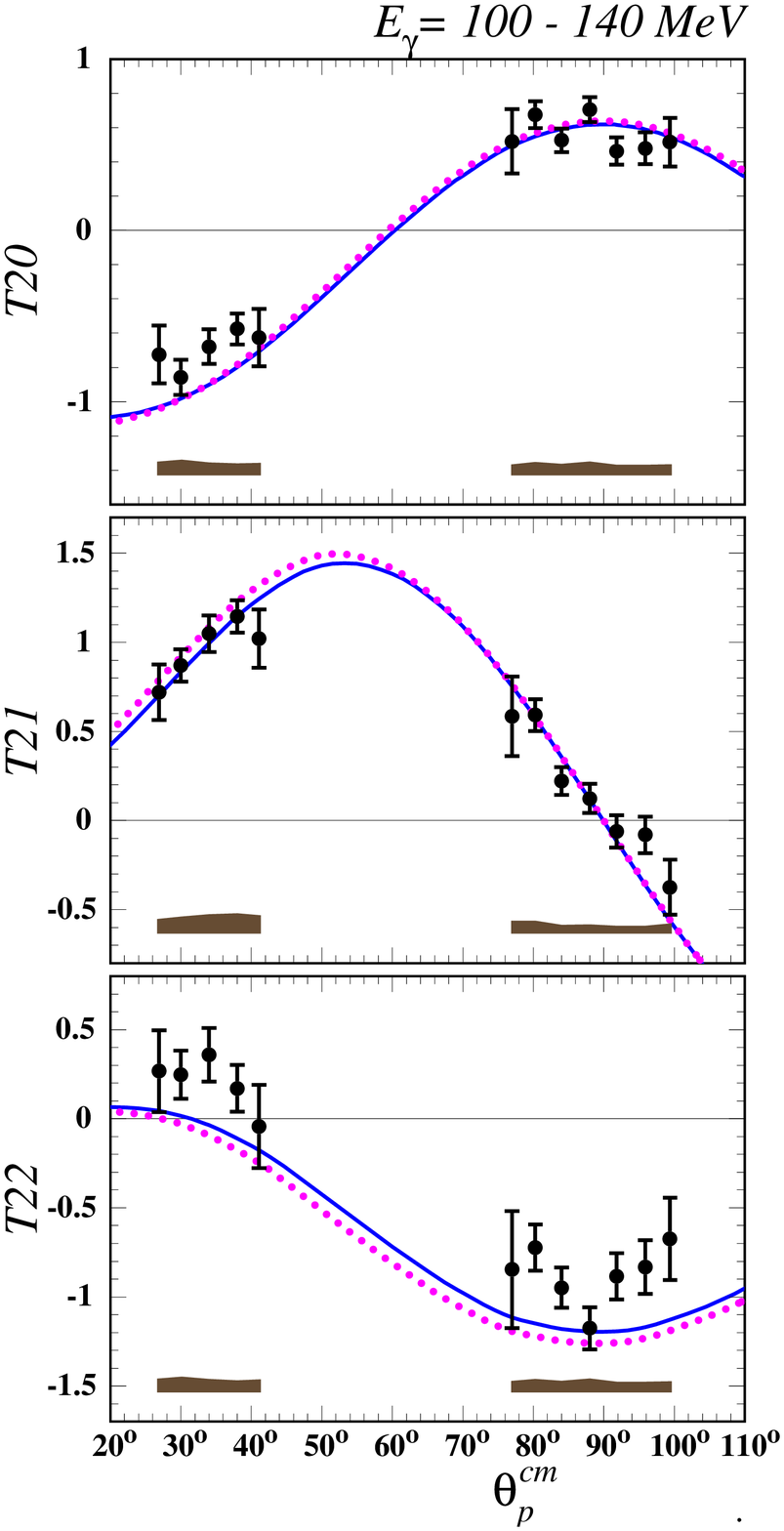}\hspace{0.01\linewidth}

\includegraphics[width=\thwid]{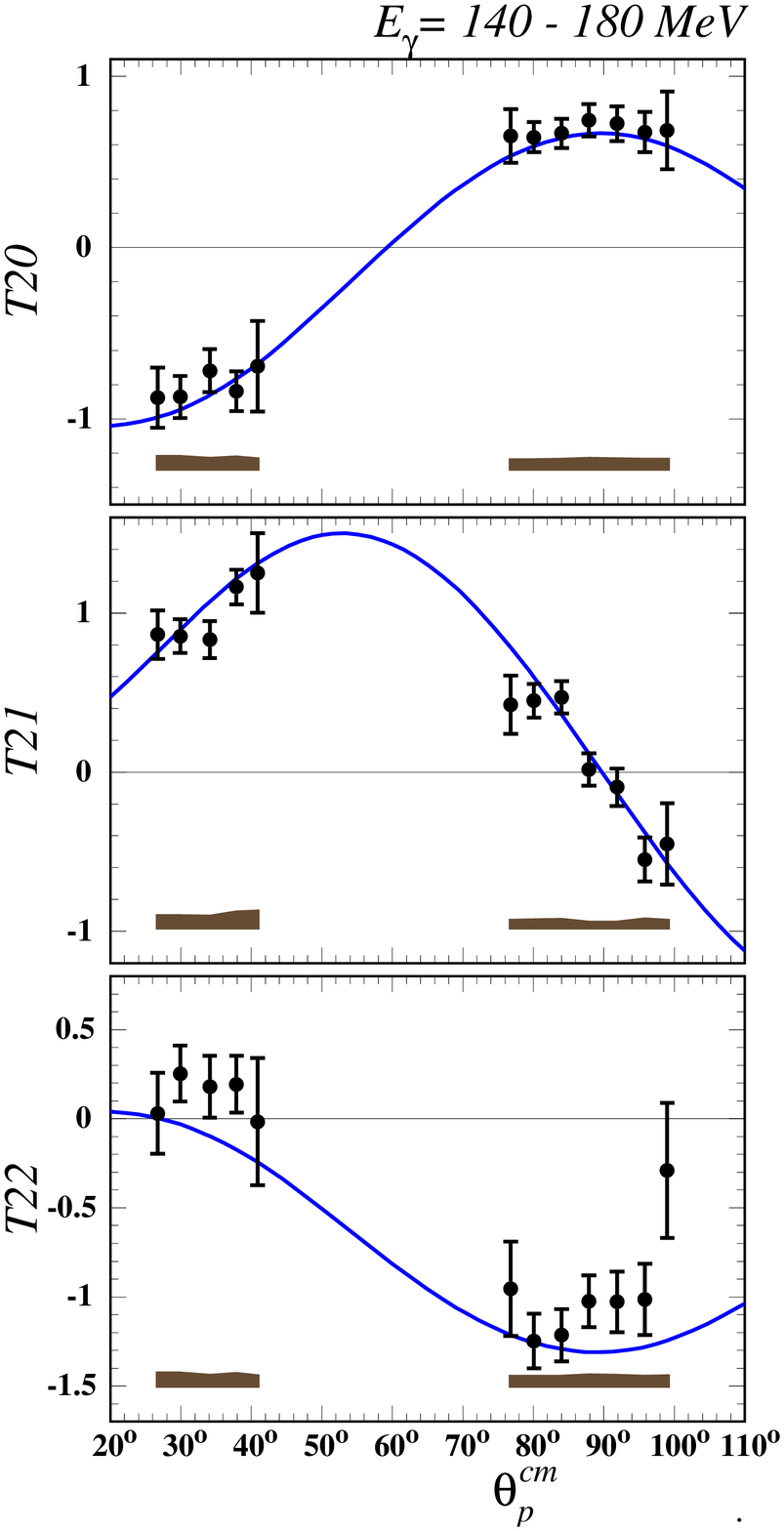}\hspace{0.01\linewidth}
\includegraphics[width=\thwid]{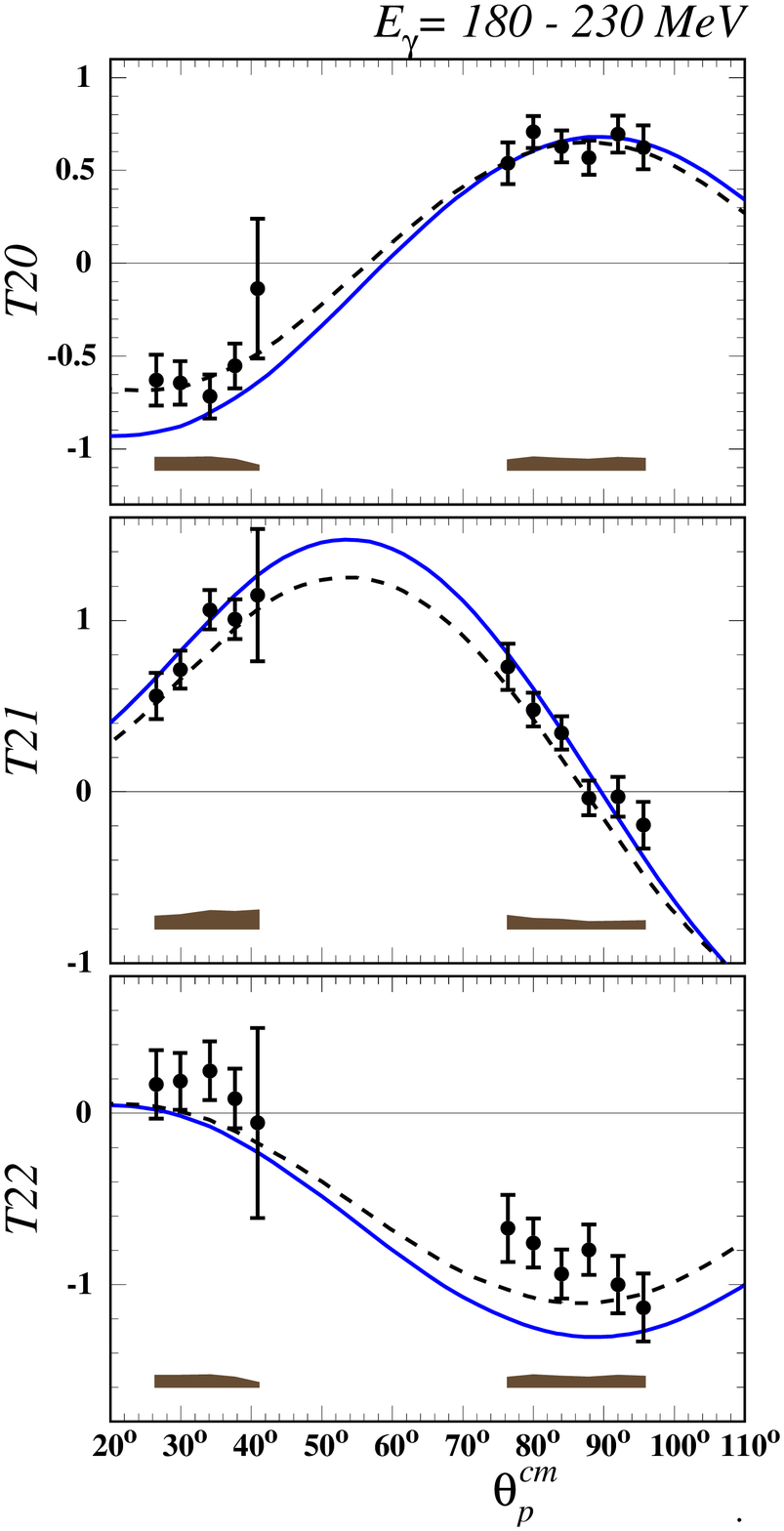}\hspace{0.01\linewidth}
\includegraphics[width=\thwid]{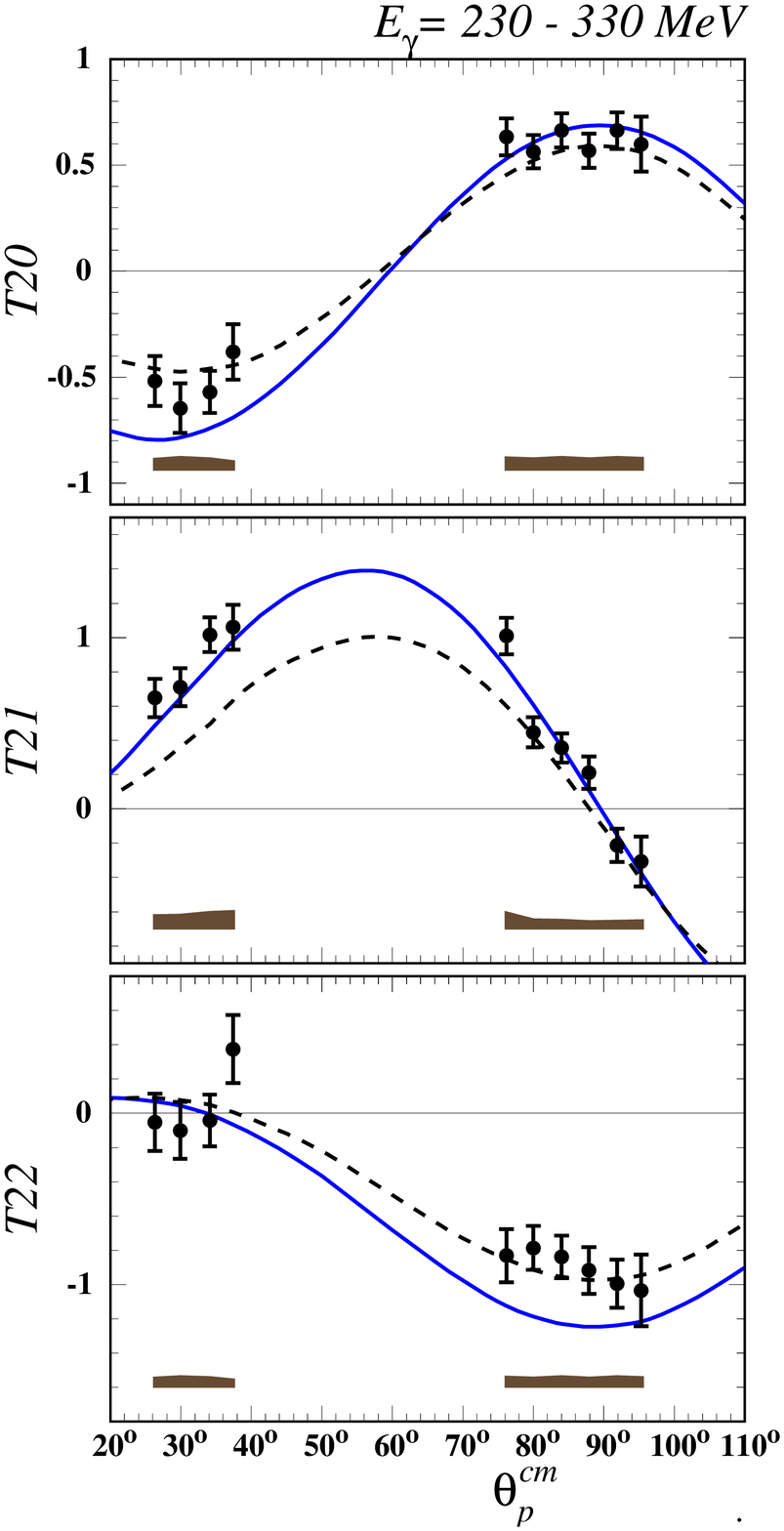}\hspace{0.01\linewidth}
\includegraphics[width=\thwid]{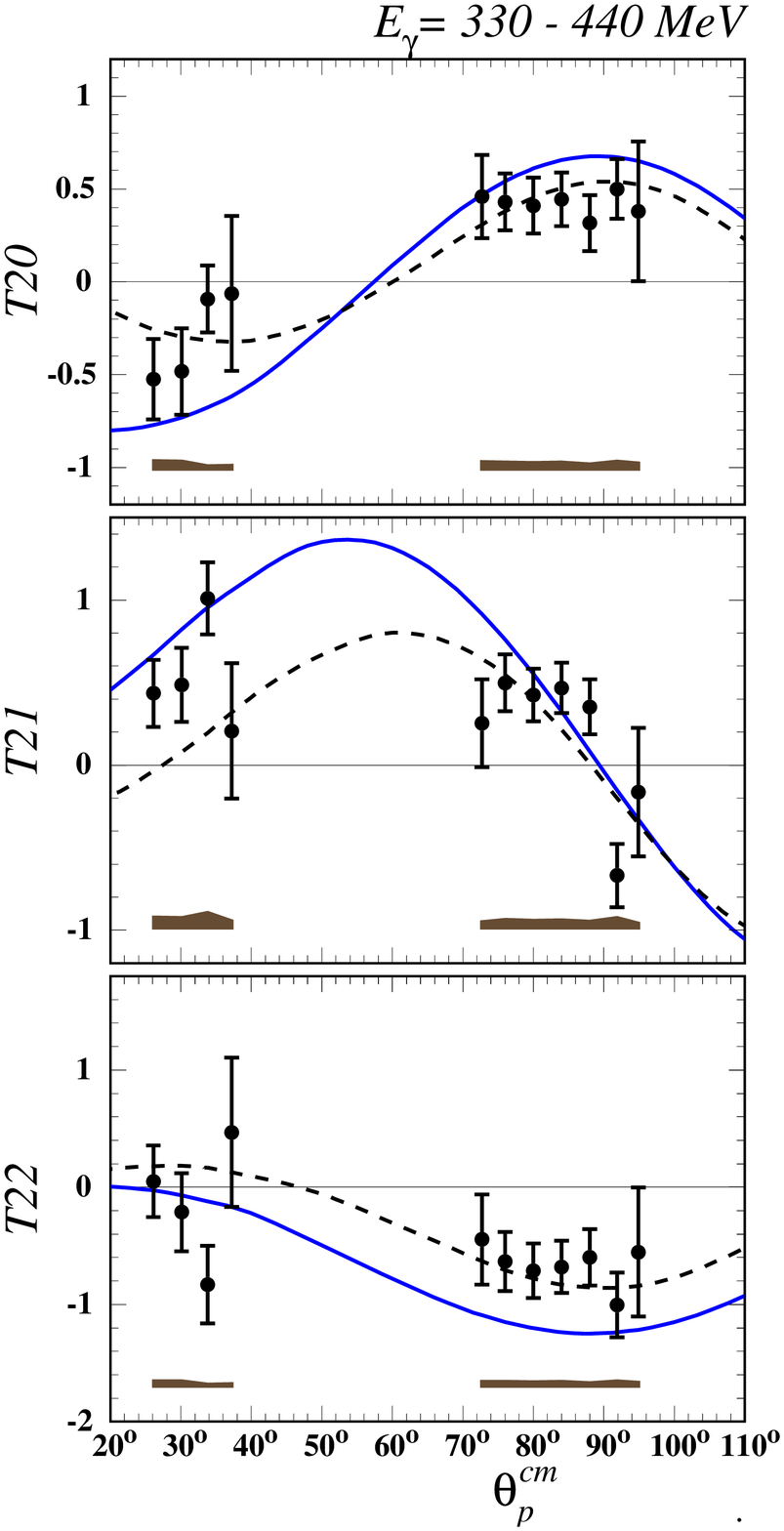}\hspace{0.01\linewidth}
\end{center}
\caption{
\label{theta}(color online)
Tensor analyzing powers vs. proton emission angle for eight $E_\gamma$-bins. 
Each $\theta_p^{cm}$ bin is $4^\circ$ wide.
See Fig.\ref{egamma} for notation.
}
\end{figure*}

In Figures \ref{egamma},\ref{theta} systematic errors are presented as shaded 
bands at top or bottom of the plots.

%\section{Results and discussion}
Tensor analyzing powers are functions of two variables, and usually 
E$_\gamma^{lab}$ and $\vartheta^{cm}_p$ are chosen. Our data cover 
substantially broad continuous regions of both variables.
Binning of the experimental data was done in order to provide both sufficient
statistical precision and a reasonable number of bins to display the
dependence of tensor moments on kinematic variables.
\begin{figure}[!tb]
\begin{center}
\includegraphics[width=0.45\linewidth]{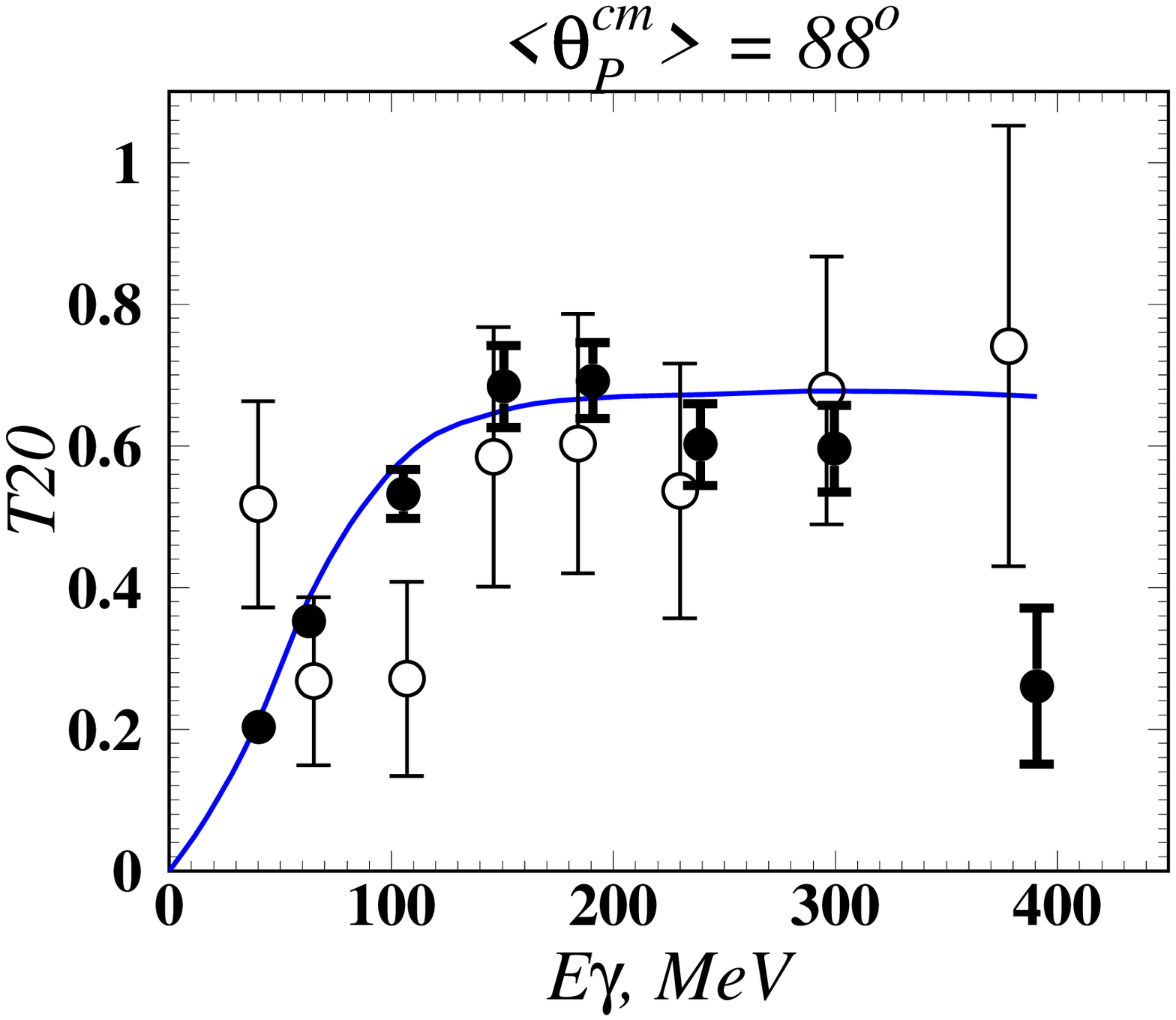}
\hspace*{0.05\linewidth}
\includegraphics[width=0.45\linewidth]{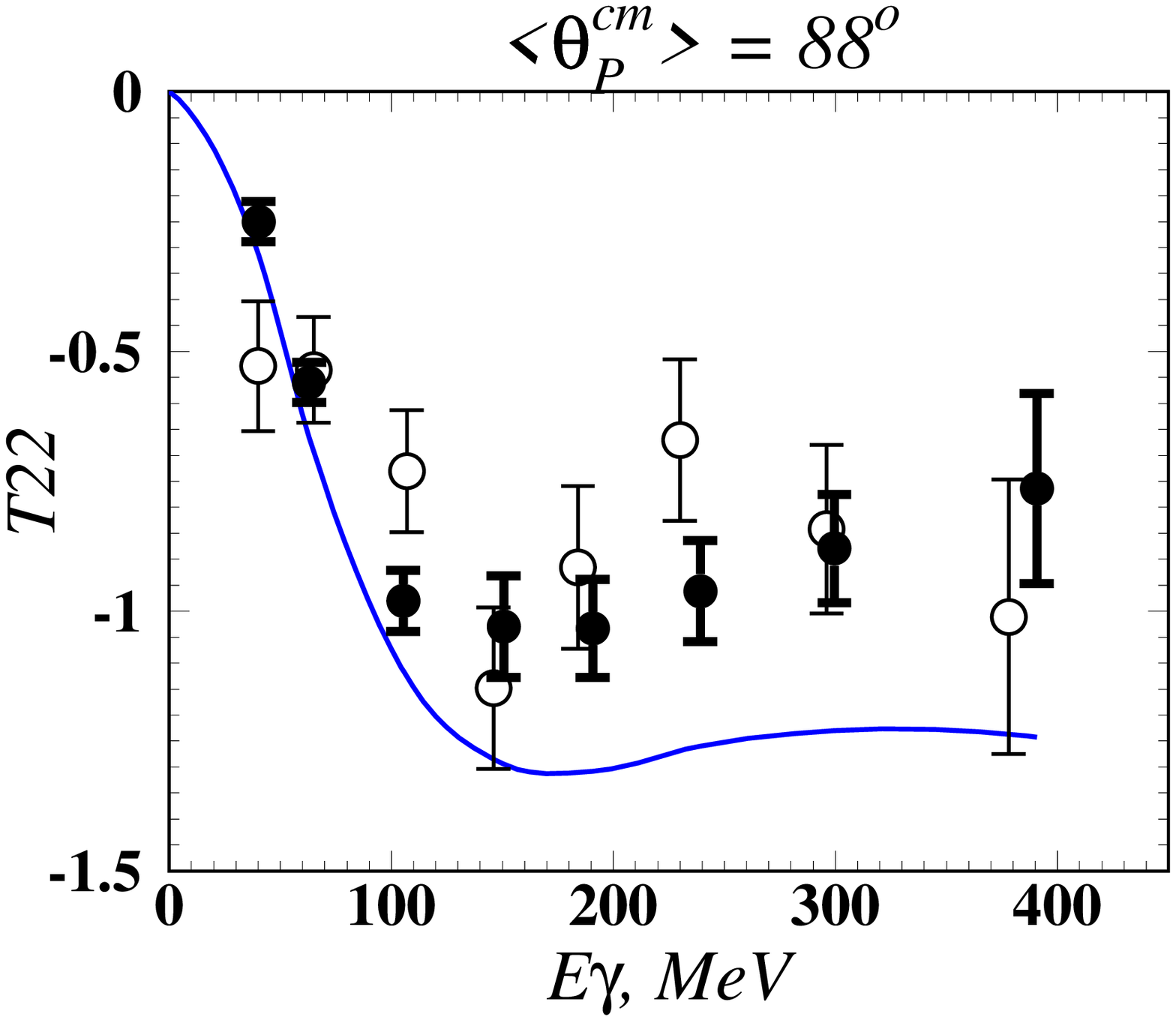}
\end{center}
\caption{
\label{oldnew}
Comparison of previous data \cite{phaseI} (open circles) and 
this work (filled circles). Only the part of the new data 
which corresponds to the kinematic conditions of the previous 
measurement is shown. Theoretical curve is
the full calculation from \cite{arenho} 
}
\end{figure}

To compare the new data with the results of earlier measurements
\cite{phaseI}, a subset of the data, where the kinematic acceptances of two
measurements overlap, was selected\footnote{Note that the data from \cite{phaseI} 
were corrected by re-analysis of the target polarimeter data, resulting in a target polarization 
P$_{zz}=0.48$ instead of P$_{zz}=0.58$ used in \cite{phaseI}.} -- see Fig.~\ref{oldnew}.
One can see that two measurements are consistent within uncertainties.

To compare the data to theoretical predictions we have performed an
event weighted averaging of the theory over the phase-space of each 
experimental bin.
The calculation \cite{arenho} starts from a one-body current using the Bonn 
OBEPR $NN$ potential with the major part of meson exchange currents (MEC) 
included implicitly via the Siegert operators; this model is denoted as 
"normal" (N). Then explicit pion exchange currents ("+MEC"), 
isobar configurations ("+IC") and the leading order relativistic corrections 
("+RC") are added successively. 
The calculation of \cite{levchuk} was done in a diagrammatic approach with 
all MEC, including heavy-meson ones, introduced explicitly, and isobar 
configurations and relativistic corrections added. 
This calculation is restricted to $E_\gamma$ below pion production 
threshold.  
In \cite{schw_are} the deuteron PD beyond pion production
threshold is studied in a coupled--channel approach including 
$N\Delta$ and $\pi d$ channels, with the dynamical treatment of the pion.
In consequence the $NN$ potential and $\pi$--MEC become retarded and 
electromagnetic loop corrections have been incorporated. In \cite{schwamb}
this concept was further elaborated and numerical results for various observables
were obtained.

The dependence of tensor analyzing powers on $E_\gamma$ is plotted in 
Fig.~\ref{egamma}. Here the whole dataset is divided into two
$\theta^{cm}_p$ bins, each related to the data from one pair of detector arms.
Alternately, in Fig~\ref{theta} the tensor moments versus $\theta^{cm}_p$ are 
shown for eight E$_\gamma$-bins.  
The numerical results are available from \cite{res}. 

%\section{discussion}

Figures~\ref{egamma} and \ref{theta} show that up to about
pion production threshold, there is little variation between the
theoretical calculations \cite{arenho,levchuk}, and there is good 
agreement between these calculations and the data.
The most noticeable difference is that $T_{22}$ tends to be slightly
more positive than calculated.
Above pion production, the calculations become significantly more 
complex due to the larger effects of relativity and the increasing
importance of additional channels, as the pion can propagate on-shell,
and the energy approaches the $\Delta$ resonance region.
We see greater variations between the older calculation of \cite{arenho}
and the more modern calculation of \cite{schwamb}.
The more modern calculations improve the description of $T_{20}$
and $T_{22}$, but do worse for $T_{21}$.
Despite the disagreement in details, it is clear that there is a good
overall qualitative description of the polarization data, which is a
difficult test for theory.
As both calculations are grounded in fits to nucleon-nucleon and 
meson photoproduction data, one would hope that the refinements 
in theory over time and the improvements in the description of the 
underlying reactions would lead to a clear improvement in all the 
deuteron PD data, but this is not the case.
The pattern of agreement between theory and experiment is
similar to that seen for other polarization observables such as $p_y$
and $\Sigma$; the quality of the agreement decreases at higher
energies.

In summary, a new measurement of tensor analyzing powers T$_{20}$,
T$_{21}$ and T$_{22}$ in deuteron photodisintegration, 
substantially enhancing the quality and kinematic span of the existing 
experimental data, has been performed. This enable an accurate test
of available models. 
Theoretical calculations provide an excellent description of these
polarization data below pion production threshold, 
while above pion production threshold a very good description 
of T$_{20}$ and T$_{22}$ 
is demonstrated by a novel approach incorporating a $\pi$-MEC retardation mechanism.
The remaining discrepancies could reflect the theoretical uncertainties or
some missing or poorly modeled underlying dynamics.

%\section*{Acknowledgments}
We gratefully acknowledge the staff of VEPP--3 accelerator
facility for excellent performance of the ring during the data taking.
We are grateful
to H.Arenh\"{o}vel, M.Levchuk and M.Schwamb for useful discussions and 
for providing us with the results of their calculations.
This work was supported in part by Russian Foundation for Basic Research,
grants 01-02-16929, 04-02-16434, 05-02-17080 and 05-02-17688,
the U.S. Department of Energy, Office of
Nuclear Physics, under contract no. W-31-109-ENG-38,
and the U.S. National Science Foundation, grant PHY-03-54871.


\begin{thebibliography}{19}

\bibitem{GG} R. Gilman and F. Gross,
J. Phys. {\bf G28}, R37 (2002).

\bibitem{vepp2}%
 M.V. Mostovoy, {\it et al.},
Phys. Lett. {\bf B189}, 181 (1987).

\bibitem{bonn}%
K.H.Althoff {\it et al.}, Z Phys. {\bf C43}, 375 (1989).

\bibitem{phaseI} S.I. Mishnev, {\it et al.},
Phys. Lett. {\bf B302}, 23 (1993).

\bibitem{ABS} M.V. Dyug {\it et al.}, Nucl. Instrum. Methods  
{\bf A495}, 8 (2002).

\bibitem{LOWQ} M.V. Dyug {\it et.al.}, Nucl. Instrum. Methods
{\bf A536}, 344 (2005). 

\bibitem{arenho}
K.-M. Schmitt and H. Arenh\"ovel, Few-Body Syst. 7, 95 (1989), 
H.Arenh\"ovel and M. Sanzone, Few-Body Syst. Suppl. 3, 1 (1991), 
F. Ritz, H. Arenh\"ovel, and T. Wilbois, Few-Body Syst. 24, 
123 (1998)

\bibitem{levchuk} M.I. Levchuk, Few-Body Syst. {\bf 19}, 77 (1995)
and private communication.

\bibitem{schw_are} M. Schwamb, H. Arenh\"ovel, Nucl. Phys. {\bf A 690}, 647
 (2001), and Nucl. Phys. {\bf A 696}, 556 (2001)
\bibitem{schwamb} M. Schwamb, habilitation thesis, Johannes Gutenberg-Universit\"at
at Mainz, 2006

\bibitem{res}http://www.inp.nsk.su/$\sim$rachek/photodisintegration.html

\end{thebibliography}
\end{document}